\documentclass[prb,twocolumn,showpacs,preprintnumbers,amsmath,amssymb]{revtex4}

\usepackage{graphicx}
\usepackage{dcolumn}
\usepackage{bm}

\begin{document}

\newcommand*{\cm}{cm$^{-1}$\,}
\newcommand*{\Tc}{T$_c$\,}


\title{High energy pseudogap and its evolution with doping in Fe-based superconductors as revealed by optical spectroscopy}

\author{N. L. Wang}
\author{W. Z. Hu}
 \altaffiliation[Present address: ]{Max Planck Department for Structural Dynamics,
 University of Hamburg, Centre for Free Electron Laser Science, Notkestra$\ss$e 85, 22607 Hamburg, Germany}
\author{Z. G. Chen}
\author{R. H. Yuan}
\author{G. Li}
\altaffiliation[Present address: ]{National High Magnetic Field
Laboratory, Florida State University, Tallahassee, Florida 32310,
USA}
\author{G. F. Chen}
 \altaffiliation[Present address: ]{Department of Physics, Renmin University of China, Beijing 100872, China}
\author{T. Xiang}

\affiliation{Beijing National Laboratory for Condensed Matter
Physics, Institute of Physics, Chinese Academy of Sciences,
Beijing 100190, China}


\begin{abstract}

We report optical spectroscopic measurements on electron- and
hole-doped BaFe2As2. We show that the compounds in the normal
state are not simple metals. The optical conductivity spectra
contain, in addition to the free carrier response at low
frequency, a temperature-dependent gap-like suppression at rather
high energy scale near 0.6 eV. This suppression evolves with the
As-Fe-As bond angle induced by electron- or hole-doping.
Furthermore, the feature becomes much weaker in the
Fe-chalcogenide compounds. We elaborate that the feature is caused
by the strong Hund's rule coupling effect between the itinerant
electrons and localized electron moment arising from the multiple
Fe 3d orbitals. Our experiments demonstrate the coexistence of
itinerant and localized electrons in iron-based compounds, which
would then lead to a more comprehensive picture about the metallic
magnetism in the materials.

\end{abstract}


\maketitle

\section{Introduction}

The interplay between superconductivity and magnetism is one of
the central topics in the study of Fe-based superconductors.
Similar to the high-T$_c$ cuprates, the superconductivity in
Fe-based compounds is found to be in close proximity to an
antiferromagnetic (AFM) order\cite{Dong,Cruz}. Superconductivity
emerges when the magnetic order was suppressed by electron or hole
doping or application of
pressure\cite{Dong,Chen1,Rotter2,Torikachvili}. Although the phase
diagram of Fe-pnictides looks very similar to that of high-$T_c$
curpates, distinct differences exist between them. The undoped
compounds in cuprates are Mott insulators, by contrast, the parent
compounds in Fe-pnictides/chalcogenides are multiband metals.
Magnetic interaction effect also manifests in the paramagnetic
phase. The uniform magnetic susceptibility $\chi$ in both parent
compounds above T$_N$ and superconducting compounds above T$_c$ is
neither Pauli- nor Curie-Weiss-like, instead it increases linearly
with increasing temperature within measured temperature
range.\cite{Zhang} Furthermore, the value of $\chi$ is roughly two
orders larger than the value for a Pauli paramagnetism. Much
debates have been focused on whether the magnetism has an
itinerant electron or local moment interaction origin, an issue
being intimately related to the pairing interaction for
superconductivity.

The complex magnetic properties arise from the multiple orbital
nature of the systems. The electronic structures of the multiple
orbital systems should be also reflected in the charge excitation
spectra which could be probed by the optical spectroscopy
technique. Previous optical investigations on the parent compounds
of AFe$_2$As$_2$ (A=Ba, Sr)\cite{Hu} revealed clearly the
formation of the partial energy gaps at low energy in the magnetic
phase, indicating the involvement of the itinerant electrons in
the spin-density-wave (SDW) order formation. Nevertheless, the gap
opening at low energy is not the sole spectral feature observed
with decreasing temperature. As indicated in the early
studies,\cite{Hu} there exists another remarkable
temperature-dependent spectral feature at much higher energy
scale. The spectral weight below $\sim$5000 \cm (0.6 eV) is
gradually suppressed with decreasing temperature and transferred
to higher energy scale (see the grey region in Fig. 1). This
suppression is observed even at room temperature. This
pseudogap-like behavior at such high energy scale is not expected
for an usual metal. Although most of the spectral weight at such
high energy scale comes from the interband transitions, the
temperature dependent part must have a different origin. It has
been a long-standing mystery for the Fe-based compounds ever since
the optical spectra were collected\cite{Hu}.

In this work, we address this issue by examining the evolution of
the high-energy pseudogap with electron- and hole-dopings. We
found that the feature remains strong in the electron-doped side,
including in the heavily electron-doped case where the compound is
no longer superconducting, but it becomes much weaker in the
heavily hole doped case. Our analysis indicates that this high
energy gap-like feature is closely related to the doping-induced
crystal structural change, particularly the As-Fe-As bond angle,
or the height of As position relative to the Fe layers. We
elaborate that the high-energy feature is caused by the strong
intra-atomic Hund's rule coupling effect due to the presence of
multiple Fe 3d orbitals. Our study supports the coexistence of
itinerant and localized electrons in iron-pnictides, which would
then lead to a more comprehensive picture about the metallic
magnetism in the materials.

\begin{figure}[t]
\includegraphics[width=6.0cm,clip]{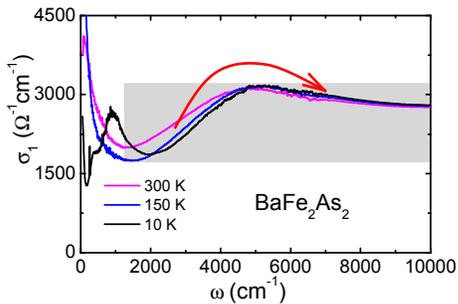}
\caption{\label{fig:R}(Color online) Optical conductivity spectra
up to 10000 \cm ($\sim$1.2 eV) for a BaFe$_2$As$_2$ single
crystal.\cite{Hu}}
\end{figure}

\section{Experiment}

Single crystals of K- and Co-doped BaFe$_2$As$_2$ were grown from
the FeAs flux method, similar to the procedure in our earlier
report.\cite{Chen2} The plate-like crystals could be easily
cleaved, resulting in very shinny surface. We present four
different samples: two K-doped Ba$_{1-x}$K$_x$Fe$_2$As$_2$ with
x=0.4 (T$_c$=37 K) and x=1 (T$_c$=3 K), and two Co-doped
BaFe$_{2-y}$Co$_y$As$_2$ with y=0.2 (T$_c$=22 K) and y=0.4 (not
superconducting). We also compare the measurement data with the
parent compounds BaFe$_2$As$_2$ and Fe$_{1.05}$Te. It is well
known that the K$^+$ doping for Ba$^{2+}$ introduces extra holes.
On the other hand, NMR experiment indicated that Co$^{2+}$ doping
for Fe$^{2+}$ does not induce local moment but offer one more
d-electron to the system, therefore has the effect of
electron-doping.\cite{Ning} Roughly, T$_c$=37 K and 22 K are close
to the highest superconducting transition temperatures achieved by
K-doing (off FeAs plane) and Co-doping (within FeAs plane),
respectively, \textit{i.e.} they are optimally doped. However,
BaFe$_{1.6}$Co$_{0.4}$As$_2$ is heavily electron-doped, and
KFe$_2$As$_2$ is heavily hole-doped. The phase diagram of the K-
and Co-doped BaFe$_2$As$_2$ is shown in Fig. 2.\cite{JHChu,Rotter}
The studied compositions are indicated in the figure.

The optical reflectance measurements were performed on a
combination of Bruker IFS 66v/s, 113v on newly cleaved surfaces
(ab-plane) of those crystals up to 25000 cm$^{-1}$. An \textit{in
situ} gold and aluminium overcoating technique was used to get the
reflectivity R($\omega$). The real part of conductivity
$\sigma_1(\omega)$ is obtained by the Kramers-Kronig
transformation of R($\omega$).

\section{Results and discussions}

\begin{figure}
\includegraphics[width=8.0cm,clip]{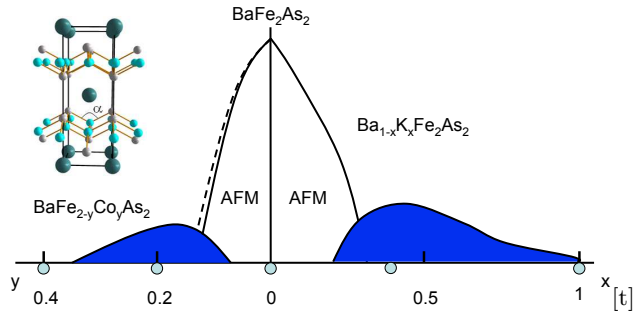}[t]
\caption{\label{fig:R}(Color online) Phase diagram of Co- and
K-doped BaFe$_2$As$_2$.\cite{JHChu,Rotter} Optical data were
presented on five different compositions as indicated in the
bottom of the phase diagram. The crystal structure of 122 compound
is shown. The As-Fe-As angle is indicated by the angle $\alpha$.}
\end{figure}

Figure 3 show the experimental reflectance and conductivity
spectra for different samples. For a comparison, we also include
the data of BaFe$_2$As$_2$ \cite{Hu} and Fe$_{1.05}$Te
\cite{Chen3} compounds. The low frequency data vary significantly
for those samples because of the different ground states.
Superconducting pairing energy gap features exist in the Co-doped
(Co=0.2) and K-doped (K=0.4) superconducting samples at very low
frequencies (not visible clearly in the plot over such broad
energy scale). Clear SDW gap features were seen for the pure
BaFe$_2$As$_2$ sample. Purely metallic temperature-dependent
responses were observed for the heavily electron- or hole-doped
samples. But here we focus our attention on the high energy
gap-like spectral weight suppression features.

\begin{figure*}
\includegraphics[width=7in]{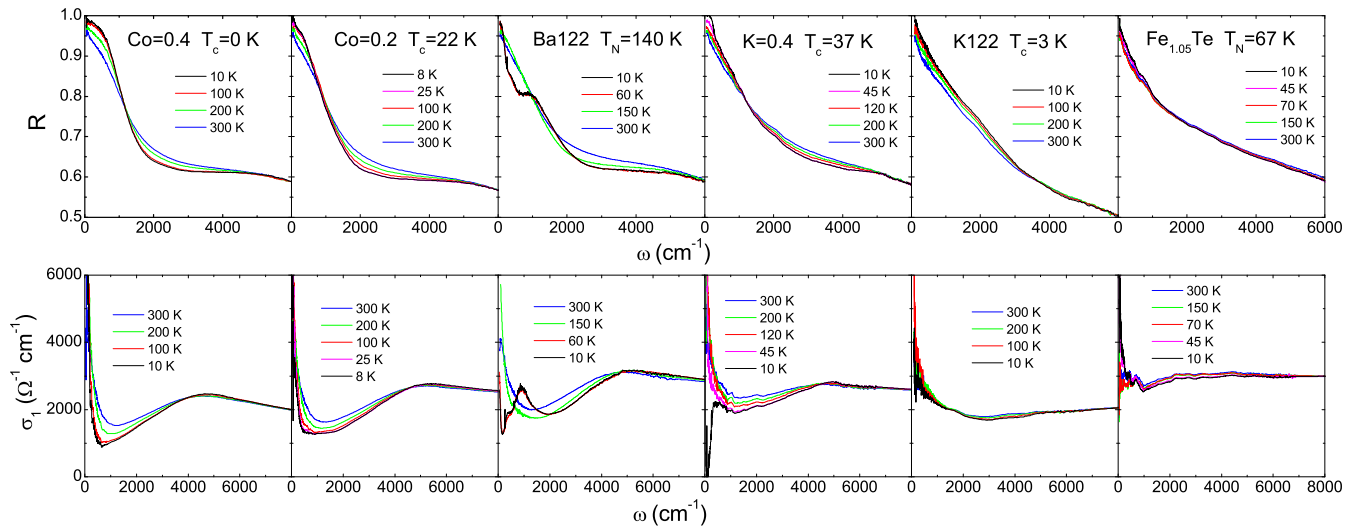}%
\vspace*{-0.20cm}%
\caption{\label{fig:R}(Color online) The evolution of the optical
spectra of BaFe$_2$As$_2$ with Co- and K-doping. Fe$_{1.05}$Te is
added for comparison. Upper panel: R($\omega$) for the pure, Co-
and K-doped BaFe$_2$As$_2$ and Fe$_{1.05}$Te up to 6000 \cm. Lower
panel: $\sigma_1(\omega)$ for for the pure, Co- and K-doped
BaFe$_2$As$_2$ and Fe$_{1.05}$Te up to 8000 \cm.}
\end{figure*}

In optical reflectance spectrum R($\omega$), the high energy
structure manifests itself as a suppression of R($\omega$) in the
mid-infrared region (about 2000-5000 \cm). As the low-$\omega$
R($\omega$) still increases towards unit due to metallic nature of
the compounds, a reverse S-like shape in R($\omega$) is resulted.
In Fig. 3 we find that, compared to the parent BaFe$_2$As$_2$
compounds, the reverse S-like shape R($\omega$) remains rather
eminent in the Co-doped compounds. However, the feature becomes
weaker in K=0.4 doped case, and tends to disappear in the pure
KFe$_2$As$_2$ sample with T$_c$ only 3 K. The feature is also not
visible in Fe$_{1.05}$Te.

Corresponding to the reverse S-like shape of R($\omega$), the real
part of conductivity $\sigma_1(\omega)$ shows a suppression
roughly below 5000 \cm. This leads to a peak in $\sigma_1(\omega)$
near this energy, as shown in Fig. 3. Our earlier study on the
parent compound indicated that the suppressed spectral weight is
transferred to higher energies,\cite{Hu} indicating a
pseudogap-like phenomenon.

\begin{figure}[b]
\includegraphics[width=8.5cm,clip]{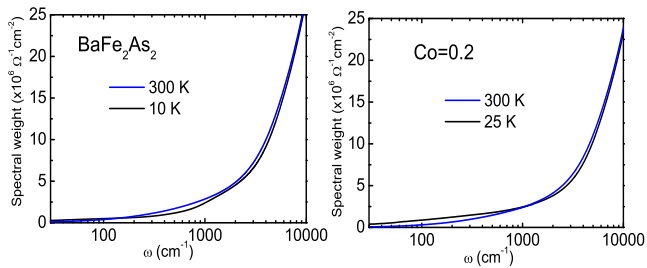}
\caption{\label{fig:R}(Color online) The integrated spectral
weight below 10000 \cm for BaFe$_2$As$_2$ and
BaFe$_{1.6}$Co$_{0.2}$As$_2$ single crystals.\cite{Hu}}
\end{figure}

To further elaborate the spectral evolution, we plot the
frequency-dependent spectral weight of $\sigma_1(\omega)$ at 10
and 300 K for two different crystals, the parent BaFe$_2$As$_2$
and Co-doped BaFe$_{1.6}$Co$_{0.2}$As$_2$, in Fig. 4. For the
parent BaFe$_2$As$_2$, a residual Drude component exists in the
SDW state.\cite{Hu} This residual Drude component narrows with
decreasing temperature, and its low-frequency limit approaches the
dc conductivity value. As a result, the low-$\omega$ spectral
weight, roughly below 200 \cm, at low T is higher than at high T.
Above 200 \cm, the SDW gap develops which strongly reduces the
low-T Drude weight, leading to the first suppression below 1000
\cm. The lost Drude weight fills into the SDW peak, and the total
spectral weight is almost recovered around 2000 \cm for 10 K.
However, the T-dependent suppression below the mid-infrared peak
results in the second spectral weight suppression at 10 K near
3000 \cm. The lost weight gradually recovers at very high energy,
roughly about 10000 \cm. For the Co-doped x=0.2 crystal, there is
no SDW gap developing at low temperature, the low-$\omega$
spectral weight change is induced by the Drude component
narrowing. A balance is seen near 1200 \cm. The further reduction
of the spectral weight beyond this frequency at low temperature is
caused by the gap-like suppression near 5000 \cm. Once again, the
recovery of the spectral weight extends to rather high energy
scale. We should remark that, at such high energy scale, most of
the spectral weight would come from the interband transitions.
However, the temperature-dependent part must have a different
origin. Seen clearly from Fig. 3, the pseudogap feature remains
very strong for all Co-doped superconducting samples, but become
weak in the K-doped compounds and tends to vanish for the pure
KFe$_2$As$_2$. No suppression feature is visible in
$\sigma_1(\omega)$ for Fe$_{1.05}$Te.

The key issue here is the origin of this high energy pseudogap.
Unlike the spin-density-wave gap observed for the parent compounds
only below magnetic ordering temperature, the high energy
pseudogap feature is present at all measurement temperatures as
well as in the doped compounds, as indicated in Fig. 3. There
could be several possibilities for the presence of high energy
peak structure. One possibility is that the quasi-particles
contain not only the coherent spectral weight at low energy but
also the incoherent part at high energies due to the presence of
strong electron correlation effect.\cite{Georges,Rozenberg,Qimiao}
The high energy feature comes from the incoherent part of the
quasi-particle spectral function. This should be a generic
phenomenon for strongly correlated electron system. A schematic
picture about the quasi-particle spectral weight function is shown
in Fig. 5. The incoherent part is argued to originate from the
onset Hubbard U. A dynamical mean-field theory calculation
indicated that both the incoherent structures at intermediate
frequencies of the order U/2 to U and the coherent Drude component
at the lower end of the conductivity spectrum rapidly emerge as
the temperature is lowered.\cite{Georges,Rozenberg} This has been
used to explain the experimental observation of development of
both Drude component and mid-infrared peak at low temperature in
V$_2$O$_3$.\cite{Rozenberg} However, this is different from the
present case where the Drude component already exists in Fe-based
compounds at high temperatures. Furthermore, this picture is
rather difficult to explain the doping evolution of the structure.
In particular, Fe$_{1.05}$Te is believed to have stronger electron
correlation, but the temperature-dependent feature at mid-infrared
is almost invisible. In fact, the conductivity spectrum in the
mid-infrared region even has slightly lower values at lower
temperature.

\begin{figure}[t]
\includegraphics[width=7cm,clip]{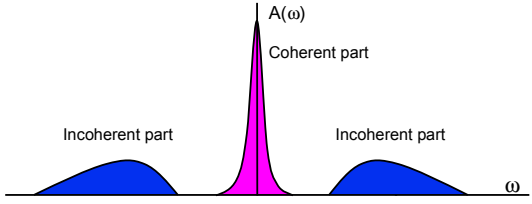}
\caption{\label{fig:R}(Color online) The spectral function of a
correlated electron system. It contains both coherent part at low
frequency and incoherent part at higher frequencies. The
incoherent part could lead to sizeable spectral weight at
mid-infrared region.}
\end{figure}

A different but related proposal is that the coherent spectrum at
low frequency is due to the itinerant electrons from some Fe 3d
orbitals which form the disconnected electron and hole Fermi
surfaces, while the incoherent part is due to the Hubbard U
splitting of the localized bands from other Fe 3d
orbitals.\cite{Kou,You} The occupation of the electrons at the
lower Hubbard band results in the formation of local moments. In
our opinion, this scenario also faces great challenges. The
electron correlations in Fe-based compounds are not so strong. The
band renormalization factors were found to be only about
2-3.\cite{Qazilbash,ZGChen,Yi,Singh1} Furthermore, some orbitals
are almost degenerate. It is hard to image that those bands could
be split by the relatively weak on-site electron correlation.

Because the feature is temperature dependent, one may think that
it is related to the spin-fluctuations. A naive idea is that the
feature is caused by the indirect interband transitions assisted
by the spin fluctuations with an AFM ($\pi, \pi$) wave vector. As
shown schematically in Fig. 6, the hole and electron bands are
well connected by such commensurate wave vector \textbf{q}=($\pi,
\pi$) in the BaFe$_2$As$_2$ parent compound. The indirect
interband transition could be realized through the assistance of
($\pi, \pi$) spin fluctuations with the transition energy of
\textit{h$\nu$}=E$_f$-E$_i$+$\Omega$, where E$_i$ and E$_f$ are
the energy levels of the initial and final states of the particle
hole excitations, $\Omega$ is the energy corresponding to the
formation of fluctuated AFM correlation which usually have a
rather small energy scale. However, a careful examination
indicates that this possibility is also unlikely. First, the bands
close to the $\Gamma$ and M points are better connected by the
($\pi, \pi$) wave vector, the lower energy region is expected to
have higher indirect interband transition spectral weight with
decreasing temperature, which is opposite to the suppression
spectral feature at lower frequencies. Second, the opposite trend
between electron- and hole-dopings highly suggests against this
scenario. The ($\pi, \pi$) spin-fluctuations are believed to be
strongly suppressed in either heavily electron- or hole-doped
compounds. Then, it is difficult to understand why they change in
the opposite ways.

\begin{figure}[t]
\includegraphics[width=7.5cm,clip]{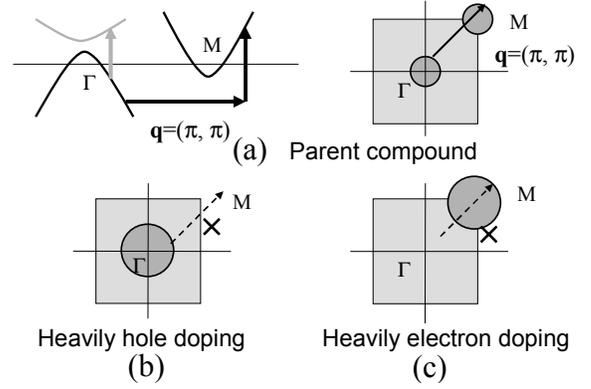}
\caption{\label{fig:R} A schematic picture for the band structure
evolution. For the parent BaFe$_2$As$_2$, indirect interband
transitions between the hole and electron bands could occur with
the help of the \textbf{q}=($\pi, \pi$) AF spin fluctuations. For
the heavily hole-doped compound, e.g. pure KFe$_2$As$_2$, the
chemical potential shifts downward, which leads to the removal of
the electron FS near M point and associated AF correlation (b).
For the heavily electron (Co-) doped case, the chemical potential
shifts up, leading to the removal of the hole FS surface near
$\Gamma$ (c). In both cases, ($\pi, \pi$) AF fluctuations are
strongly suppressed. The spin fluctuation-assisted indirect
interband transition should be significantly reduced.}
\end{figure}

We propose that the high energy feature reflects the presence of
local physics. The spectral weigh transfer from low energy to high
energy implies a transfer of relatively itinerant electrons with
binding energy less than 0.6 eV to the more localized part. A
promising picture is as follows. The Fe-pnictides/chalcogenides
contain both itinerant electrons and local moments arising from
different 3d orbitals of the iron atoms. Unlike the proposals
mentioned above, the local moment formation is not caused by the
Hubbard U interaction but mainly originate from the Hund's rule
coupling interaction of the different orbitals. If the Hund's rule
coupling energy is smaller than the kinetic energy (or band width)
of electrons in some orbitals, those electrons are itinerant. On
the other hand, if the Hund's coupling energy is larger than the
kinetic energy of the electrons in other orbitals, those electrons
are localized and form local moments. In general, the itinerant
electrons and local moments are not completely isolated, but are
also coupled via the Hund's rule interaction, as illustrated in
Fig. 7. We identify the energy scale of the suppression near 0.6
eV as the local Hund's rule coupling energy scale. This value is
indeed close to those estimated in a number of the theoretical
works\cite{Haule,GTWang,Yin,Moon,Johannes} and a resonant
inelastic x-ray scattering (RIXS) experiment \cite{Yang}. At low
temperature, the thermal excitations become weak, then a fraction
of relatively itinerant electrons tend to become more localized.
Thus, we would expect to observe the spectral weight transfer from
low energy to an energy scale higher than the Hund's rule coupling
energy.

The above picture provides a natural explanation for the
experimental observation in the optical measurement. Nevertheless,
there are two crucial questions which have to be addressed. First,
why a relatively small change of temperature from 300 K to 10 K
could lead to such an apparent spectral weight transfer over very
broad frequency range, that is, from far below 0.6 eV to far above
it? Second, why the spectral weight transfer feature evolves
significantly with electron or hole doping?

\begin{figure}
\includegraphics[width=7.5cm,clip]{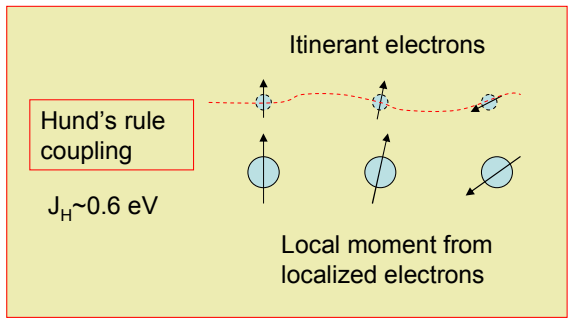}
\caption{\label{fig:R} A schematic picture about the Hund's
coupling between itinerant electrons and local moments. The local
moment tends to polarize the spin of the itinerant electrons in a
ferromagnetic orientation. At low T, itinerant electrons tends to
become more localized due to this coupling, leading to the
spectral weigh transfer from low energy to higher energies.}
\end{figure}

To understand the first question, we have to assume that the
kinetic energies of the electrons in different Fe 3d
bands/orbitals are relatively close. Some may be slightly larger
than the Hund's rule coupling energy, while others may be slightly
smaller than the Hund's rule coupling energy. In other words, the
energy levels or bands that contribute to the local moments are
rather close to the Fermi level where itinerant electrons
dominate, so that the temperature change could affect their
interaction. Indeed, the first principle band structure
calculations on LaFeAsO and Ba122 indicated that all 5 orbitals of
Fe 3d electrons are rather close in energy.\cite{Singh1} The local
moments estimated from the density function calculations are
usually higher than $ 2\mu_B$, suggesting more than 2 electrons
are localized which contribute to the local moments. The situation
is somewhat similar to the element $\alpha$-Fe, which has been
known for several decades for the presence of both itinerant and
localized d electrons. It deserves to remark that the recent
neutron \cite{Zhijun,Zaliznyak} and RIXS experiments revealed
presence of large local moments in both parent compounds above
T$_N$ and doped superconducting compounds above T$_c$.

An important observation in the present work is that the spectral
weight transfer feature changes with doping, particularly in the
K-doped case. From the crystal structural characterization, it is
found that the K-doping results in a continuous decrease of the
a-axis but an increase of the c-axis parameters. More detailed
analysis indicates that K-doping does not induce detectable change
in the Fe-As bonding length, but lead to a continuous decrease of
As-Fe-As angle (the $\alpha$ angle in the crystal structure shown
in Fig. 2) \cite{Rotter}. Equivalently, the height of As atom
relative to the Fe-layer is increasing. The As height in FeTe is
further increased compared to Fe-pnictides.\cite{Yin} On the
contrary, the Co-doping does not lead to an detectable change in
the a-axis lattice parameter but only a slight decrease of c-axis
parameter \cite{CTLin}. According to the band structural
calculations, the band structure is rather sensitive to the As
height relative to the Fe layer.\cite{GTWang,Yin,Moon}  The
increase of As height mainly make the Fe atom more isolated,
leading to the narrowing the Fe 3d bands. As a result, the
effective Hund's rule coupling is enhanced.\cite{Yin} Then the
spectral weight suppression feature could not be seen within a
temperature range change from 300 to 10 K. Higher temperature
range would be required in order to see the spectral weight
transfer structure. We found that the picture could well explain
the doing evolution and the disappearance of the feature in
Fe$_{1.05}$Te.

Finally, we comment on the spin-density-wave gap formation on the
Fe-pnictide compounds. Based on the above study, the local moment
formation itself is not due to the Fermi surface nesting, but
caused by the strong Hund's rule coupling between different Fe 3d
orbitals. However, the Fermi surfaces couple strongly with the
magnetic instability. A spin-density-wave gap would open once the
nesting of FS's matches with the magnetic wave vector.

\section{Summary}

In conclusion, our measurement indicates that the FeAs compounds
are not simple itinerant electron systems. The optical data of the
122 parent compound and their evolution with K- and Co-doping
revealed spectral change not only at low but also considerable
high frequencies. The spectral structure and its evolution implies
the presence of both itinerant and localized electrons arising
from different Fe 3d orbitals. We elaborate that the spectral
weight transfer over the broad energies is related to the Hund's
rule coupling between itinerant and localized electrons. The
coupling effect could be strongly affected by the environment of
Fe atom, i.e. the bonding angle of As-Fe-As or the height of As
atom relative to the Fe layer. Our experiments demonstrate the
coexistence of itinerant and localized electrons in iron-based
compounds, which would then lead to a more comprehensive picture
about the metallic magnetism in the materials.

\begin{acknowledgments}
We would like to thank Z. Y. Weng, T. Egami, W. Ku, W. G. Yin, H.
J. Choi, G. M. Zhang, J. P. Hu, J. L. Luo, D. H. Lee, Z. Fang, X.
Dai for helpful discussions. This work is supported by the NSFC,
CAS, and the 973 project of the MOST of China.
\end{acknowledgments}


\end{document}